\documentclass[a4paper,12pt]{article}
\usepackage[english]{babel}
\usepackage{amssymb,amsmath,amsfonts,amsthm}
\usepackage{hyphenat}
\usepackage{latexsym}
\usepackage{graphicx}
\usepackage{feynmf}

\newcommand{\be}{\begin{equation}}
\newcommand{\ee}{\end{equation}}
\newcommand{\bd}{\begin{definition}}
\newcommand{\ed}{\end{definition}}
\newcommand{\bt}{\begin{theorem}}
\newcommand{\et}{\end{theorem}}
\newcommand{\bp}{\begin{proof}}
\newcommand{\ep}{\end{proof}}
\newcommand{\bea}{\begin{eqnarray}}
\newcommand{\eea}{\end{eqnarray}}
\newcommand{\ba}{\begin{array}}
\newcommand{\ea}{\end{array}}

\def\siml{{\ \lower-1.2pt\vbox{\hbox{\rlap{$<$}\lower6pt\vbox{\hbox{$\sim$}}}}\ }}

\def\al{\alpha}

\newcommand{\nn}{\nonumber}

\newcommand{\pref}{2m_{p}\left[\frac{eg_{A}}{2f_{\pi}}\right]^{2}}

\begin{document}

\begin{titlepage}
\begin{flushright}
\end{flushright}

\vspace{1cm}
\begin{center}
\begin{Large}
{\bf Forward virtual Compton scattering and the Lamb shift 
in chiral perturbation theory}\\[2cm] 
\end{Large} 
{\large David Nevado$^{1}$ and Antonio Pineda$^{2}$}\\
\vspace{0.5cm}
{\it $^{1}$ Dept. d'Estructura i Constituents de la Mat\`eria and IFAE,
U. Barcelona \\ Diagonal 647, E-08028 Barcelona, Spain}\\
\vspace{0.5cm}
{\it $^{2}$ Grup de F\'\i sica 
Te\`orica and IFAE, Universitat Aut\`onoma de Barcelona,\\ 
E-08193 Bellaterra (Barcelona), Spain}
\end{center}

\vspace{1cm}

\begin{abstract}
We compute the spin-independent 
structure functions of the forward virtual-photon Compton 
tensor of the proton at one loop using heavy baryon chiral  
perturbation theory and dispersion relations. We study 
 the relation between both approaches. We use these results 
to generalize some sum rules to virtual photon transfer momentum and 
relate them with sum rules in deep inelastic scattering. 
We then compute the leading chiral 
term of the polarizability correction to the 
 Lamb shift of the hydrogen and 
muonic hydrogen. We obtain $-{87.05}/{n^3}\,{\rm Hz}$ and $-{0.148}/{n^3}\,{\rm meV}$ 
for the correction to the hydrogen and muonic hydrogen Lamb shift respectively. 
\vspace{5mm} \\
\end{abstract}

\end{titlepage}
\vfill
\setcounter{footnote}{0} 
\vspace{1cm}

\section{Introduction}

The knowledge of the forward virtual-photon Compton tensor of the proton, $T^{\mu\nu}$, 
for energies of the 
order of the pion mass allows one to check chiral perturbation theory 
in the (but not deep) Euclidean region ($Q^2 \equiv -q^2 \sim m_{\pi}^2 \not= 0$). 
This is interesting by itself, since it allows us to test whether chiral perturbation 
theory computations work equally well in the Euclidean or Minkowski regime. 
Moreover, the expressions obtained for the structure functions permit one to generalize 
sum rules typically derived for real photons ($q^2=0$) to virtual photons 
in the Euclidean regime ($q^2 \sim -m_{\pi}^2 <0$). 
These results provide us with a more direct connection with perturbation theory 
calculations, which are valid for the very same Euclidean sum rules but at large $Q^2$. This idea has 
already been addressed in the case of the spin-dependent structure 
functions in Ref. \cite{Ji:1999mr}. It is the main motivation of this paper 
to obtain the spin-independent generalized sum rules. To do 
so we will compute the spin-independent structure functions at one loop
in heavy baryon chiral perturbation theory \cite{Gasser:1983yg,HBET}.

These results are also a necessary step in the complete determination of the 
leading chiral-related polarizability corrections to the Lamb shift of the hydrogen and 
muonic hydrogen\footnote{In the same way that the spin-dependent structure 
functions are necessary for the hyperfine splitting of the 
hydrogen and muonic hydrogen \cite{Pineda:2002as}.}. 
These corrections are model independent and may yield    
accurate determinations of the proton radius provided the muonic 
hydrogen Lamb shift is measured with high enough accuracy \cite{Antognini:2005fe}. So far 
only some partial results have been obtained \cite{Pineda:2004mx}. 
In this paper we fill this gap by obtaining the complete expression for the 
leading correction of the polarizabity effect to the Lamb shift. 

We will then compute $T^{\mu\nu}$ using Feynman rules and study its 
relation with dispersion relation computations in depth.
We will find this computation to be a useful exercise to illustrate the problems 
and advantages of working with time-ordered correlators 
versus its non-ordered counterparts. We will clearly see 
that the derivation 
of correct dispersion relations may turn out to be more difficult than expected,
and, in particular, that the seagull terms are necessary in order to restore 
current conservation.

The expressions for the structure functions of $T^{\mu\nu}$ can be understood as 
generalized sum rules for $Q^2 \not= 0$. By expanding around $\rho \equiv q\cdot p/m_p=0$, 
we can relate our results with the sum rules in the deep Euclidean region. We will  
discuss the connection between the chiral and perturbative regime but a 
quantitative analysis will be relegated to future work.

The structure of the paper is the following. In sec. 2 we define the tensors and 
form factors we use in our computation. In sec. 3 we compute $W^{\mu\nu}$ and 
$T^{\mu\nu}$ using Feynman rules. In sec. 4 we compute $T^{\mu\nu}$ 
using dispersion relations and compare with the previous result. 
In sec. 5 we analyze the resulting sum rules. In sec. 6 we compute the 
polarizability corrections to the Lamb shift. 
In sec. 7 we give our conclusions. In the Appendix some contributions 
to $T^{\mu\nu}$ due to the $\Delta$ particle are computed. 

\section{Definitions}

We first define the form factors, which we will understand 
as pure hadronic quantities, i.e. without electromagnetic 
corrections. 

We define $J^\mu=\sum_i Q_i{\bar q}_i\gamma^\mu q_i$ where $i=u,d$ (we will 
not consider the strange quark in this paper)
and $Q_i$ is the quark charge. The 
form factors are then defined by the following equation:   
\be
\langle {p^\prime,s}|J^\mu|{p,s}\rangle
=
\bar u(p^\prime) \left[ F_1(q^2) \gamma^\mu +
i F_2(q^2){\sigma^{\mu \nu} q_\nu\over 2 m_p} \right]u(p)
\label{current}
\,,
\ee
where $q=p'-p$  and $F_1$, $F_2$ are the Dirac and Pauli form factors,
respectively. The states
are normalized in the following (standard relativistic) way:
\be
\langle p',\lambda'|p,\lambda \rangle
=
(2\pi)^3 2p^0\delta^3({\bf p}'-{\bf p})
\delta_{\lambda' \lambda}\,,
\ee
and 
\be
u(p,s){\bar u}(p,s)=({\rlap/p+m_p }){1+\gamma_5{\rlap/s} \over 2}
\,,
\ee
where $s$ is an arbitrary spin four vector obeying $s^2=-1$ and
$p\cdot s=0$.
 
The form factors could be (analytically) expanded as follows 
\be
F_i(q^2)=F_i+{q^2 \over m_p^2}F_i^{\prime}+...\,
\ee 
for very low momentum. Nevertheless, we will be interested instead in their 
(non-analytic) behavior in $q \sim m_{\pi}$. 

We also introduce the Sachs form factors:
\be
G_E(q^2)=F_1(q^2)+{q^2 \over 4m_p^2}F_2(q^2), \qquad G_M(q^2)=F_1(q^2)+F_2(q^2). 
\ee

We also define
\begin{equation}  \label{forw-amp2}
 W^{\mu\nu} = {1 \over 4\pi}\!\int\! d^4x\, e^{iq\cdot x}
  \langle {p,s}| [ J^\mu(x),J^\nu(0)] |{p,s}\rangle
\,,
\end{equation}
which has the following structure ($\rho=q\cdot p/m_p$ and we 
take the sign convention $\epsilon^{0123}=1$):
\bea \label{Wmunu}
 W^{\mu\nu} &=
  &\left( -g^{\mu\nu} + \frac{q^\mu q^\nu}{q^2}\right) W_1(\rho,q^2) 
  + \frac1{m_p^2} \left( p^\mu - \frac{m_p\rho}{q^2} q^\mu \right)
    \left( p^\nu - \frac{m_p\rho}{q^2} q^\nu \right) W_2(\rho,q^2)  \\*
	\nn
  && - \frac i{m_p}\, \epsilon^{\mu\nu\rho\sigma} q_\rho s_\sigma G_1(\rho,q^2)
  - \frac i{m_p^3}\, \epsilon^{\mu\nu\rho\sigma} q_\rho
   \bigl( (m_p\rho) s_\sigma - (q\cdot s) p_\sigma \bigr)
G_2(\rho,q^2)=W_S^{\mu\nu}+W_A^{\mu\nu}
\,,
\eea
where $W_S^{\mu\nu}=W_S^{\nu\mu}$ and $W_A^{\mu\nu}=-W_A^{\nu\mu}$. 
Therefore, the first two (spin-independent) terms correspond to $W_S^{\mu\nu}$ 
and the last two (spin-dependent) terms correspond to $W_A^{\mu\nu}$.

We will also need the forward virtual-photon Compton tensor
\begin{equation} 
 T^{\mu\nu} = i\!\int\! d^4x\, e^{iq\cdot x}
  \langle {p,s}| T \{J^\mu(x)J^\nu(0)\} |{p,s}\rangle
\,,
\end{equation}
which has the following structure ($\rho=q\cdot p/m_p\equiv v \cdot q$ although we will usually work in the 
rest frame where $\rho=q^0$):
\bea \label{inv-dec}
 T^{\mu\nu} &=
  &\left( -g^{\mu\nu} + \frac{q^\mu q^\nu}{q^2}\right) S_1(\rho,q^2) 
  + \frac1{m_p^2} \left( p^\mu - \frac{m_p\rho}{q^2} q^\mu \right)
    \left( p^\nu - \frac{m_p\rho}{q^2} q^\nu \right) S_2(\rho,q^2) \nn 
	\\*
  && - \frac i{m_p}\, \epsilon^{\mu\nu\rho\sigma} q_\rho s_\sigma A_1(\rho,q^2)
	- \frac i{m_p^3}\, \epsilon^{\mu\nu\rho\sigma} q_\rho
   \bigl( (m_p\rho) s_\sigma - (q\cdot s) p_\sigma \bigr) A_2(\rho,q^2),
\eea
depending on four scalar functions, which we call structure functions. 
Similarly to $W^{\mu\nu}$, we can split the tensor in the symmetric,
$T_S^{\mu\nu}=T_S^{\nu\mu}$, and antisymmetric piece, $T_A^{\mu\nu}=-T_A^{\nu\mu}$. 
Moreover, it is also usual to split these 
tensors in two components, which we label $Born$ and $pol.$, for instance,
$T^{\mu\nu}=T^{\mu\nu}_{Born}+T^{\mu\nu}_{pol.}$. 
Each of these terms complies with current conservation\footnote{The separation 
(definition) of the $Born$ and $pol.$ terms is in general ambiguous, see Refs. 
\cite{Scherer:1996ux,Fearing:1996gs} for an extensive discussion on this point. 
In our case, as far as we give an explicit definition 
for $T^{\mu\nu}_{Born}$, this ambiguity disappears.}. 
The Born term is the contribution coming from the intermediate state being the proton (somewhat the 
elastic contribution) and the associated structure functions can be written in terms of the form factors. 
They read
\bea
 S_1^{\rm Born}(\rho,q^2) & =
  -2F_1^2(q^2) - \frac {2(q^2)^2\,G_{\rm M}^2(q^2)}{(2m_p\rho)^2-(q^2)^2}, \\
 S_2^{\rm Born}(\rho,q^2) & =
  2\, \frac {4m_p^2q^2\,F_1^2(q^2)-(q^2)^2\,F_2^2(q^2)}{(2m_p\rho)^2-(q^2)^2}, \\
 A_1^{\rm Born}(\rho,q^2) & =  \label{Born.A1}
  -F_2^2(q^2) + \frac {4m_p^2q^2\,F_1(q^2)G_{\rm M}(q^2)}{(2m_p\rho)^2-(q^2)^2}, \\
 A_2^{\rm Born}(\rho,q^2) & =
  \frac {4m_p^3\rho\,F_2(q^2)G_{\rm M}(q^2)}{(2m_p\rho)^2-(q^2)^2}.
\eea
From these expressions one can easily single out the point-like contributions (those due to 
a point-like particle but with non-trivial anomalous dimension, radius, ...). The remaining contributions,  
with the one-loop accuracy of our chiral computation, are encoded in the 
following expression (we split $G_{E,M}$ into pieces according to its chiral counting:
$G_{E,M}^{(n)} \sim 1/m_p^n\sim 1/\Lambda_{\chi}^n$):
\bea
\label{TBorn}
&&T^{\mu\nu}_{Born}
=
i\pi\delta(v\cdot q)
\\
&&
\nn
\times
Tr\left[u\bar u
\left(
-4p_+G_E^{(0)}G_E^{(2)}v^{\mu}v^{\nu}
+\frac{2}{m_p}G_E^{(0)}G_M^{(1)}
\left(
v^{\mu}p_+\left[s^{\nu},s^{\rho'}\right]q_{\rho'}p_+
-
v^{\nu}p_+\left[s^{\mu},s^{\rho'}\right]q_{\rho'}p_+
\right)
\right)
\right]
\,,
\eea
where $p_+=\frac{1+v\cdot \gamma}{2}$. 
Note that $T^{\mu\nu}_{Born}$ is proportional to $\delta(v\cdot q)$. $G_E^{(0)}=1$. The expressions for $G_E^{(2)}$, $G_M^{(1)}$ can be found in 
Refs. \cite{Bernard:1992qa,BKKM,BFHM}\footnote{In Ref. \cite{Pineda:2002as} $G_M^{(1)}$ was labeled $G_M^{(2)}$, and is meant to express 
one loop in chiral perturbation theory.}. For the spin-dependent case, the only contribution is the term proportional to 
$G_E^{(0)}G_M^{(1)}$, which comes from the $A_1^{\rm Born}$ term (this is the only term that contributes to the Zemach contribution 
to the hyperfine splitting). 
For the spin-independent case, on which we focus in this paper, we would only need $G_E^{(2)}$.

Note that some point-like contributions are still encoded in $G_E^{(2)}$ and $G_M^{(1)}$. 
Nevertheless, it should also be stressed that, in many situations, only the non-analytic 
behavior in the momentum of the structure functions is 
really relevant. In those cases it is not necessary to obtain 
the complete expressions for the structure functions, and, in particular,  
counterterms can be avoided. This comment applies to 
the computation of the non-analytic behavior in the light quark masses  
and in the splitting between nucleon and the Delta mass 
(proportional to powers of $1/N_c$ in the large $N_c$ limit) of the Lamb shift 
(or hyperfine splitting) from the Sachs form factors. This is because only the non-analytic behavior of the structure function 
gives a non-zero contribution to the dimensional regularized 
one-loop computation. 

In what follows we consider the computation of $T^{\mu\nu}_{S,pol.}$, which is novel.

\section{Computation of $T^{\mu\nu}$ and $W ^{\mu\nu}$}

We will restrict our analysis to the SU(2) flavour case  
and neglect the $\Delta$ particle. 
We can obtain a compact expression for the polarizability 
contribution to $W^{\mu\nu}$ 
at tree level (for simplicity we set $v^{\mu}=(1,0)$):
\bea
\label{Wpol}
W_{S,pol.}^{\mu\nu}&=&\pref\int \frac{d^3{\bf p}_{\pi}}{2E_{\pi}(2\pi)^3}\delta(E_{\pi}-q^0)
\\
&&
\times
\left\{g^{\mu i}-\frac{g^{\mu 0}\:p^{i}_{\pi}}{q^0}+\frac{k^{i}\:\left(q-2k\right)^{\mu}}
{k^2-m_{\pi}^2+i\eta}\right\}\left\{g^{\nu i}-\frac{g^{\nu 0}\:p^{i}_{\pi}}{q^0}+
\frac{k^{i}\:\left(q-2k\right)^{\nu}}{k^2-m_{\pi}^2+i\eta}\right\},\nonumber
\eea
where $E_{\pi}=\sqrt{{\bf p}_{\pi}^2+m^2_{\pi}}$, $k=(0,{\bf k})=(0,{\bf q-p_{\pi}})$, $k=q-p_{\pi}$.

It is easy to check that the above expression satisfies current conservation. From this result one 
could try to obtain $W_1$ and $W_2$ directly, instead we will obtain them later from the 
imaginary part of $S_i$.

\begin{figure}[h]
	\centering
		\includegraphics[width=10cm]{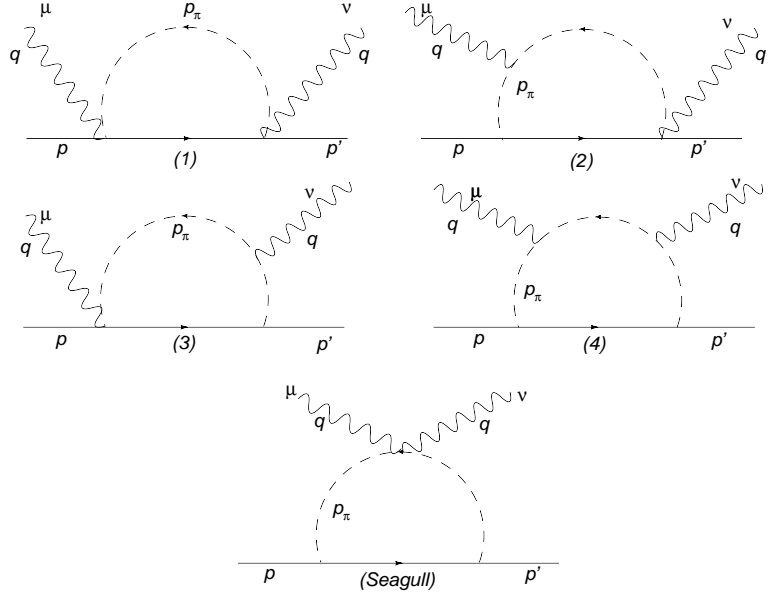}
\caption{\label{figTij}Diagrams contributing to $T^{ij}$. Crossed diagrams are not 
explicitly shown but calculated. 
}
\end{figure}

The expressions we obtain for $S_1$ and $S_2$ using chiral perturbation theory read
\bea
\label{S1pol}
{S}_{1}^{pol.}&=&\frac{1}{\pi}\:\left(\frac{g_{A}}{2f_{\pi}}\right)^2\:m_{p}\:m_{\pi}\:\left\{\frac{3}{2}+\frac{m_{\pi}^2}{{\bf q}^2}-\left(1+\frac{m_{\pi}^2}{{\bf q}^2}\right)\:\sqrt{1-z}\right.\\
&-&\left.\frac{1}{2}\:\sqrt{\frac{m_{\pi}^2}{{\bf q}^2}}\:\left(2+\frac{q^2}{{\bf q}^2}\right)\:\mathcal{I}_{1}\,(m_{\pi}^2,q^{0},q^2)\right\}\nonumber
\,,
\eea
where
\be
z=\frac{(q^{0})^2}{m_{\pi}^2}
\ee
and
\bea\label{eq:integraldef1}
\mathcal{I}_{1}\,(m_{\pi}^2,q^{0},q^2)
&=&\:\int_{0}^{1}dx\frac{1}{\sqrt{\frac{m_{\pi}^{2}}{{\bf q}^{2}}-\frac{q^{2}}{{\bf q}{2}}\:x-x^{2}}}
\\&=&
-\arctan\left(\frac{q^2}{2m_{\pi}|{\bf q}|}\right)
+\arctan\left(\frac{2{\bf q}^2+q^2}{2|{\bf q}|\sqrt{m_{\pi}^2-q_0^2}}\right)
\nn
\eea
\bea
\label{S2pol}
{S}_{2}^{pol.}&=&\frac{1}{\pi}\:\left(\frac{g_{A}}{2f_{\pi}}\right)^2\:m_{p}\:m_{\pi}\:
\frac{q^2}{{\bf q}^2}
\left\{-\left(\frac{3}{2}+\left(\frac{1}{2}+\frac{m_{\pi}^{2}}{q^{2}}
+\frac{m_{\pi}^{2}}{\left(q^{0}\right)^{2}}\right)\frac{q^{2}}{{\bf q}^{2}}\right)\right.\nonumber\\
&-&\frac{\left(q^{0}\right)^{2}\:q^{2}}{4m_{\pi}^{2}{\bf q}^{2}+(q^{2})^{2}}\:\left(\frac{m_{\pi}^2}{{\bf q}^2}-\frac{{ q}^2}{2{\bf q}^2}\right)\\
&+&\frac{m_{\pi}^{2}}{{\bf q}^{2}}\left(2-\frac{{\bf q}^{2}}{ \left(q^{0}\right)^{2}}\left(1-z\right)+\frac{q^{2}\:\left(q^{0}\right)^{2}}{4m_{\pi}^{2}{\bf q}^{2}+(q^{2})^{2}}\right)\:\sqrt{1-z}\nonumber\\
&+&\left.\frac{1}{2}\sqrt{\frac{m_{\pi}^{2}}{{\bf q}^{2}}}\:\left(2+3\frac{q^{2}}{{\bf q}^{2}}+\frac{q^{2}}{m_{\pi}^{2}}\right)\:\mathcal{I}_{1}\,(m_{\pi}^2,q^{0},q^2)\right\}\nonumber
\,.
\eea
We have obtained these results by computing the spin independent part of 
$T^{ij}=A\delta^{ij}+Bq^iq^j$ from the diagrams shown in Fig. \ref{figTij}. 
We note that only inelastic diagrams contribute to $T^{ij}$. 
$ T^{00}$ has 
been obtained afterwards using current conservation\footnote{We have actually checked that 
this relation also holds by a direct computation of $T^{00}$ using Feynman diagrams.} 
\be
\label{currcons}
T^{00}=\frac{q^iq^jT^{ij}}{q_0^2}
\,.
\ee
Note that this relation may miss singular terms (delta-type) in $q^0$. 
Those are precisely the ones that appear in the Born term (actually current 
conservation holds for both the born and polarizability corrections 
independently). Polarizability terms should not have $\delta(q^0)$ terms. 
From $T^{00}$ and $T^{\mu}_{\mu}$, we can then reconstruct $S_i^{pol.}$ 
obtaining the results above. 

\section{Dispersion relations}

We now try to reconstruct $T^{\mu\nu}$ from  $W^{\mu\nu}$ 
using dispersion relations. In principle, this procedure could be  
convenient in chiral computations, since it reduces the number of diagrams to 
be considered, because $W^{\mu\nu}$ can be obtained with one order less. Nevertheless, 
in our case this difference is not significant. 

The structure functions of $W^{\mu\nu}$ are the ones that appear in
$ep$ inelastic scattering. We can relate them to the structure functions
 of $T^{\mu\nu}$ by
using its spectral decomposition, which relates the imaginary term of $T ^{\mu\nu}$
to $W^{\mu \nu}$:
\be
W^{\mu\nu}(q)|_{q^0>0}={1 \over 2\pi} {\rm Im} T^{\mu\nu}(q)|_{q^0>0}
\,.
\ee
We can use this equality to obtain $W_1$ and $W_2$. They read
\bea
\nonumber
W_{1}^{pol.}(q^0,q^2)
&=&
{1 \over 2\pi} {\rm Im} S_1^{pol.}
=
\theta(q_0^2-m_{\pi}^2)
\frac{1}{2\pi^2}
\:\left(\frac{g_{A}}{2f_{\pi}}\right)^2
\:m_{p}\:m_{\pi}
\:\left\{
\left(1+\frac{m_{\pi}^2}{{\bf q}^2}\right)\:\sqrt{q_0^2-m_{\pi}^2}
\right.\\
&-&
\left.\frac{1}{4}
\:\sqrt{\frac{m_{\pi}^2}{{\bf q}^2}}
\:\left(2+\frac{q^2}{{\bf q}^2}\right)
\:\ln\left[\frac{q_0^2+{\bf q}^2+\sqrt{4{\bf q}^2(q_0^2-m_{\pi}^2)}}
{q_0^2+{\bf q}^2-\sqrt{4{\bf q}^2(q_0^2-m_{\pi}^2)}}\right]
\right\}
\,,
\label{ImW1}
\\
\label{ImW2}
W_{2}^{pol.}(q^0,q^2)&=&
{1 \over 2\pi} {\rm Im} S_2^{pol.}
=
\theta(q_0^2-m_{\pi}^2)
\frac{1}{2\pi^2}
\:\left(\frac{g_{A}}{2f_{\pi}}\right)^2
\:m_{p}\:m_{\pi}
\frac{q^2}{{\bf q}^2}
\\
\nn
&&
\times
\:\left\{
-\frac{m_{\pi}^2}{{\bf q}^2}
\left(2-\frac{{\bf q}^{2}}{ q_{0}^{2}}\left(1-
\frac{q_0^2}{m_{\pi}^2}\right)
+\frac{q^{2}\:q_{0}^{2}}{4m_{\pi}^{2}{\bf q}^{2}+(q^{2})^{2}}\right)
\:\sqrt{q_0^2-m_{\pi}^2}
\right.\\
&+&\left.\frac{1}{4}\sqrt{\frac{m_{\pi}^{2}}{{\bf q}^{2}}}\:\left(2+3\frac{q^{2}}{{\bf q}^{2}}+\frac{q^{2}}{m_{\pi}^{2}}\right)
\:\ln\left[\frac{q_0^2+{\bf q}^2+\sqrt{4{\bf q}^2(q_0^2-m_{\pi}^2)}}
{q_0^2+{\bf q}^2-\sqrt{4{\bf q}^2(q_0^2-m_{\pi}^2)}}\right]
\right\}\nonumber
\,.
\eea

On the other hand, by using the spectral decomposition (naively, without caring about convergence), we obtain the following 
equalities\footnote{The expressions for $T^{\mu\nu}$ obtained from Feynman rules and dispersion relations 
may differ. Therefore, we label those obtained with dispersion relations with the label $DR$.}
\be
T_{S,DR}^{0i}(q)={1\over 2}(T_{DR}^{0i}(q)+T_{DR}^{i0}(q))
=
4\int_{q_{0,th}}^{\infty}
d q_0^{\prime} q_0 {W_S^{0i}(q_0^{'},{\bf q}^2) \over q_0^{'2}- q_0^{2}-i\epsilon}
\,,
\ee
and
\be
\label{Tmn}
T_{S,DR}^{\mu\nu}(q)={1\over 2}(T_{DR}^{\mu\nu}(q)+T_{DR}^{\nu\mu}(q))
=
2\int_{q_{0,th}}^{\infty}
d q_0^{'2} {W_S^{\mu\nu}(q_0^{'},{\bf q}^2) \over q_0^{'2}- q_0^{2}-i\epsilon}
\,,
\ee
otherwise. It should be stressed that the equalities are derived at ${\bf q}^2=$ $constant$. 
This is quite different from the standard dispersion relations that are derived 
at $Q^2=-q^2=$ $constant$. The latter are derived using the Cauchy theorem and 
the assumed analytic properties of the structure functions. In principle, the same results should 
be obtained in both approaches (up to, maybe, some polynomials in $q^0$). 

Using Eq. (\ref{Tmn}), the first four (plus the crossed ones) diagrams in fig. \ref{figTij} 
can be obtained using dispersion relation techniques on a diagram-by-diagram basis. If we 
want to obtain the complete result using dispersion techniques the last 
diagram in fig. \ref{figTij} is missing. This could be expected because, in general, the use of dispersion relations 
reproduces the full result up to, may be, some local terms, 
polynomials in $q^0$, which are usually called seagull terms. 
The last 
diagram can be obtained by direct computation (being a constant, it can also be 
fixed from the normalization constant of $A$ (which is momentum independent) 
in the limit of real photons and $q^0 \rightarrow 0$, since this limit 
is known.). Overall we find
\be
T_S^{ij}=T_{S,DR}^{ij}+T_{seagull}^{ij}
\,,
\ee
where
\be
T_{seagull}^{\mu\nu}
=
-g^{\mu\nu}\frac{3}{2\pi}\:\left(\frac{g_{A}}{2f_{\pi}}\right)^2\:m_{p}\:m_{\pi}
\,.
\ee

For the "00" component we find (we do not consider $\delta(q^0)$ contributions, 
which are encoded in the Born term)
\be
T_S^{00}=T_{S,DR}^{00}
\,.
\ee
In this case there is not a one-to-one correspondence between the diagrams obtained 
from Feynman rules and the dispersion relation analysis. 
Note also that current conservation is lost if we only consider the result obtained 
from the dispersion analysis. The seagull term is needed for the "ij" component, 
whereas for the "00" component the seagull term cancels the no one-to-one correspondence 
between the Feynman diagrams and the dispersion relation computation.

\section{Generalized Sum Rules}

We have checked that in the limit of real photons ($q^2=0$) our expressions 
for $S_i^{pol.}$ agree with those obtained in Ref. \cite{Bernard:1992qa}.  Obviously in the 
 limit of $q^0=|{\bf q}| \rightarrow 0$, we also obtain the polarizabilities, $\alpha_E$ and 
 $\beta_E$, which  correspond to some specific sum rules. 

We can now take the opposite limit: $\rho \rightarrow 0$ with $Q^2$ constant.
This limit allows us a closer connection with the sum rules that appear in 
deep inelastic scattering. If we define the moments of deep inelastic sum rules as 
(we recall that $x=Q^2/(2p\cdot q)$)
\be
\label{MN}
M_N^{(i)}(Q^2) = \int_0^1 dx x^{N-1} W_i(x,Q^2) 
\,,
\ee
then the structure functions read (due to the 
symmetry of the structure functions only moments with even $N$ 
appear in our case)
\be
S_i(\rho,Q^2)=\sum_{N=0,2,4,...}\rho^NS_i^{(N)}(Q^2)
\,,
\ee
where 
\be
\label{SiN}
S_i^{(N)}(Q^2)=4\int_{Q^2/(2m_p)}^{\infty}\frac{d\rho}{\rho^{N+1}}W_i(\rho,Q^2)
=
\left(\frac{2m_p}{Q^2}\right)^NM_N^{(i)}(Q^2)
\,.
\ee

Let us note that in the standard derivation of Eq. (\ref{SiN}) one assumes that 
one can neglect the behavior at infinity. In our case we have an expression for 
$S_i(\rho,Q^2)$, which we can Taylor expand around $\rho=0$ to obtain expressions 
for $M_N^{(i)}(Q^2)$ at $Q^2 \sim m_{\pi}^2$. Therefore, $S_i(\rho,Q^2)$ by itself 
can be understood as a generalized sum rule, where the behavior 
at $\rho \sim Q^2 \sim m_{\pi}^2$ of $S_i(\rho,Q^2)$ is dictated by the 
integral of $W_i(\rho',Q^2)$ over all values of $\rho'$ weighted by the function $1/(\rho^{\prime 2}-\rho^2)$. 
Actually, it is argued in Ref. \cite{Ji:1999mr} that, in general, it is more convenient to 
directly work with $S_i(\rho,Q^2)$ rather than with the moments because of the 
existence of the inelastic thresholds. Either way our predictions can be tested 
with experiment through dispersion relations. Note as well that this is not necessarily 
 equivalent to checking with experiment their 
imaginary parts at low energies, because the sum rules involve the integration 
over an arbitrary large momentum. Therefore, the naive use of Eq. (\ref{SiN}) with the 
chiral expressions obtained in Eqs. (\ref{ImW1}) and (\ref{ImW2}) for $W_i$ 
may yield wrong results. 
This is because of the not fast enough convergence of $W_i(\rho,Q^2)$ at infinity, which 
does not allow us  to neglect its contribution there (nevertheless 
one can do so if enough derivatives are taken, but then one has to fix the 
low energy unknown constants).

Eq. (\ref{MN}) allows us to relate in a single quantity perturbation theory and the chiral 
expansion. At large $Q^2$, one can use perturbation theory to know their scaling 
with $Q^2$ (for a recent state of the art see \cite{Vermaseren:2005qc}, for instance). 
At low $Q^2$ one can use the chiral expressions obtained here (the combination of the 
elastic and inelastic terms). 

Note that in Eq. (\ref{SiN}) both the elastic and inelastic contribution have 
to be added by definition. At large $Q^2$ the dominant contribution comes from the 
inelastic term. Therefore, this one should match with the perturbative behavior at 
large $Q^2$. On the other hand, our computation holds at small $Q^2 \sim m_{\pi}^2 \ll \Lambda_{\chi}$. 
In this situation the largest contribution may come from the elastic term. 
Either way one may consider sum rules for elastic or inelastic terms and 
compare them with the chiral predictions. One example could be the Baldin sum rule, 
which corresponds to the sum of the polarizabilities: 
$\alpha_E+\beta_E$, and can be obtained from dispersion relations. 
The generalized Baldin sum rule would correspond (up to normalization) to 
\be
M_2^{(1)}(Q^2)-({\rm elastic \; term})
\,.
\ee
For a discussion on the generalized Baldin sum rule see, for 
instance, Refs. \cite{Drechsel:2002ar,Liang:2004tk}. In this paper we stop here, 
and relegate the quantitative analysis of the sum rules, including their known behavior 
at low (chiral) and high (perturbation theory) energies, as well as a 
comparison with experiment, to future work. 

\section{Lamb shift}

An important application of our result is the evaluation of the chiral-related 
polarizability effects to the Lamb shift of the hydrogen and muonic hydrogen. 
As we have already mentioned in the introduction, 
this result is of special importance in the case of the muonic hydrogen, since
the measurement of the muonic hydrogen Lamb shift may yield a very precise 
determination of the proton radius. At this respect polarizability effects appear 
as one of the main sources of uncertainty. Therefore, a precise (and model independent)
determination of their value can significantly pin down the errors of the proton radius. 
Here we provide this value for the purely chiral-related effects. 

We refer to \cite{Pineda:2004mx} (see also \cite{Pineda:2002as}) 
for further details, here we only quote the main formulas we need 
for the computation. In that reference, the polarizability effects were encoded in a matching coefficient
named $c_{3,NR}^{pl_i}$. The expression for $c_{3,NR}^{pl_i}$ can be passed to the Euclidean, which then reads
\bea
\label{c3}
&&
c_{3,NR}^{pl_i}=- e^4 m_pm_{l_i}\int {d^4k_E \over (2\pi)^4}{1 \over k_E^4}{1 \over
k_E^4+4m_{l_i}^2k_{0,E}^2 }
\\
\nn
&&
\times
\left\{
(3k_{0,E}^2+{\bf k}^2)S_1(ik_{0,E},-k_E^2)-{\bf k}^2S_2(ik_{0,E},-k_E^2)
\right\}
\,,
\eea
where $m_{l_i}$ is the mass of the lepton.

Using the explicit expressions for $S_{i}^{pol.}$ obtained in 
Eqs. (\ref{S1pol}) and (\ref{S2pol}) one could 
obtain the chiral contribution to $c_{3,NR}^{pl_i}$ due to the polarizability 
effects. In particular, it is convenient to rewrite $c_{3,NR}^{pl_i}$ 
in the following manner, which is amenable 
for a numerical analysis:
\bea
\label{c3num}
&&
c_{3,NR}^{pl_i}=- e^4 m_p^2\frac{m_{l_i}}{m_{\pi}}\left(\frac{g_A}{f_{\pi}}\right)^2
\int {d^{D-1}k_E \over (2\pi)^{D-1}}{1 \over (1+{\bf k}^2)^4}
\int_0^{\infty}\frac{dw}{\pi}w^{D-5}{1 \over
w^2+4\frac{m^2_{l_i}}{m_{\pi}^2}\frac{1}{(1+{\bf k}^2)^2}}
\nn
\\
&&
\times
\left\{
(2+(1+{\bf k}^2)^2)A_E(w^2,{\bf k}^2)+(1+{\bf k}^2)^2{\bf k}^2w^2B_E(w^2,{\bf k}^2)
\right\}
\,,
\eea
where (for $D=4$)
\bea
&&
A_E=-\frac{1}{4\pi}
\left[-\frac{3}{2}+\sqrt{1+w^2}
+\int_0^1dx\frac{1-x}{\sqrt{1+x^2w^2+x(1-x)w^2{\bf k}^2}}
\right]
\,,
\eea
\bea
\nn
&&
B_E=\frac{1}{8\pi}
\left[
\int_0^1dx\frac{1-2x}{\sqrt{1+x^2w^2+x(1-x)w^2{\bf k}^2}}
-
\frac{1}{2}\int_0^1dx\frac{(1-x)(1-2x)^2}{(1+x^2w^2+x(1-x)w^2{\bf k}^2)^{\frac{3}{2}}}
\right]
\,.
\eea

The functions $A_E$ and $B_E$ correspond to the Euclidean version of the 
functions $A$ and $B$ that appear in $T^{ij}=A\delta^{ij}+Bk^ik^j$ 
up to a normalization factor and some momentum rescaling. 

Note that Eq. (\ref{c3num}) is finite. Nevertheless, one has to be careful in 
the computation, since the contributions proportional to $A_E$ and $B_E$ are not 
finite by themselves and a cancellation of infinities needs to take place between 
both terms.

The limit of small lepton mass 
agrees (with logarithmic accuracy) with the result of Ref. \cite{Pineda:2004mx} in Eq. (51), which 
could be rewritten in terms of the polarizabilities \cite{Friar:1997tr,KS}.

Following again Ref. \cite{Pineda:2004mx}, the contribution to the potential coming from the polarizability correction reads 
\be
\delta V = -{ c_{3,NR}^{pl_i} \over m_p^2}\delta^{(3)}({\bf r})
\,,
\ee
and its contribution to the lamb shift reads ($\langle \rangle$ stands for the average over polarizations)
\be
\delta E(n)=\langle E(n,l=0)-E(n,l=1) \rangle =-{ c_{3,NR}^{pl_i} \over m_p^2}\delta_{l0}{1 \over \pi}
\left(
{\mu_{l_ip}\al \over n}
\right)^3
\,,
\ee
where $\mu_{l_ip}=m_{l_i}m_p/(m_{l_i}+m_p)$.

We can now put some numbers. For the case of Hydrogen we obtain  
\be
\left.
\delta E_{e p}^{\rm pol.}\right|_{\Delta 
\rightarrow \infty}= -\frac{87.0488}{n^3}\,{\rm Hz}
\,.
\ee
This number should be compared with the number obtained in the logarithmic approximation 
in Ref. \cite{Pineda:2004mx}:
$$
\left.
\delta E_{e p,log.}^{\rm pol.}\right|_{\Delta 
\rightarrow \infty}= -\frac{64.4841}{n^3}\,{\rm Hz}
\,.
$$
We can see that the logarithmic approximation works reasonably well. This could 
be expected since the ratio $m_e/m_{\pi}$ is a small number, yet the finite piece is sizeable. 
In any case, at present, the polarizability correction is beyond the experimental accuracy of the hydrogen Lamb shift 
due to the $m_e/m_{\pi}$ suppression factor. 

For the case of the muonic hydrogen we have $m_{\mu}/m_{\pi}$, which is not suppressed. Numerically, 
we obtain
\be
\left.\delta E_{\mu p}^{\rm pol.}\right|_{\Delta 
\rightarrow \infty}= -\frac{0.147614}{n^3}\,{\rm meV}
\,.
\ee
In this case the logarithmic approximation does not work. On the other hand, 
for the muon case, the approximation in which the mass of the muon is equal to the 
pion mass in $c_{3,NR}^{pl_i}$ is a good approximation:
\be
\left.\delta E_{\mu p}^{\rm pol.}\right|_{\Delta 
\rightarrow \infty}(m_{\mu}=m_{\pi})= -\frac{0.132339}{n^3}\,{\rm meV}
\,.
\ee
If we compare these numbers with the Zemach and vacuum polarization
hadronic contributions obtained in Ref. \cite{Pineda:2004mx}, we can 
see that they are more or less of the same size. This should be 
contrasted with the hyperfine case \cite{Pineda:2002as}, 
where the Zemach correction was the largest contribution 
(in this last case there is not vacuum polarization correction).

Our numbers are quite similar (within the expected uncertainties) to those obtained from models 
or from dispersion relations with experimental data for the 
form factors \cite{KS,pachucki2,Rosenfelder:1999px,Faustov:1999ga,Martynenko:2005rc}.

Finally, let us stress that our results give the leading term (in the chiral counting) 
to the Zemach and polarizability correction. Note as well that our result is parameter free, 
no new counterterm is needed. Therefore, it is model independent. The correction 
to our expression is $O(m_{\pi}/\Lambda_{QCD})$ suppressed. The scale of $\Lambda_{QCD}$ is 
typically dictated by the next resonance that has not been integrated out, in 
our case the $\Delta$, or, being more specific, the proton-$\Delta$ mass difference: 
$m_{\Delta}-m_p$. This produces the largest uncertainty to our result. We expect to 
compute these corrections elsewhere. 
 
\section{Conclusions}

We have computed the spin-independent 
structure functions of the forward virtual-photon Compton 
tensor of the proton at one loop using heavy baryon chiral  
perturbation theory and dispersion relations. We have studied 
the relation between both approaches, and used our results 
to generalize some sum rules to virtual photon transfer momentum and 
relate them with sum rules in deep inelastic scattering. 
We have then computed the leading chiral 
corrections to the polarizability effect of the 
 Lamb shift of the hydrogen and 
muonic hydrogen. We have obtained $-{87.05}/{n^3}\,{\rm Hz}$ and $-{0.148}/{n^3}\,{\rm meV}$ 
for the correction to the hydrogen and muonic hydrogen Lamb shift respectively. 

Extensions of this work include the computation of the $\Delta$ effects, as well as repeating 
the computation in SU(3) (including the strange quark). On the other hand we 
expect to perform a quantitative application of these results to the muonic hydrogen, 
as well as to the generalized sum rules.

\medskip

{\bf Acknowledgments} \\ 
A.P. acknowledges discussions with Santi Peris and Juan Rojo. 
This work is partially supported by the 
network Flavianet MRTN-CT-2006-035482, by the spanish 
grant FPA2007-60275, and by the catalan grant SGR2005-00916.

\appendix

\section{$\Delta$ contribution}

It is not the aim of this paper to obtain the full set of $\Delta$-related 
contributions to the structure functions, which we relegate to future 
work, yet we can not avoid displaying a few terms (those which appear at tree level)
that are easy to obtain.

For the elastic $\Delta$-dependent terms we obtain
\bea
W_1(q_0,q^2)&=&{4 \over 9} \left({b_{1,F} \over 2m_p}\right)^2
m_p\delta(q_0-\Delta){\bf k}^2
\,,
\nn
\\
W_2(q_0,q^2)&=&-{4 \over 9} \left({b_{1,F} \over 2m_p}\right)^2
m_p\delta(q_0-\Delta)k^2
\,.
\eea

\bea
S_1(q_0,q^2)&=&{16 \over 9} \left({b_{1,F} \over 2m_p}\right)^2
{\bf q}^2{m_p \Delta \over \Delta^2-q_0^2-i\epsilon} 
\,,
\nn
\\
S_2(q_0,q^2)&=&{16 \over 9} \left({b_{1,F} \over 2m_p}\right)^2
  (-\Delta^2+{\bf q}^2){ m_p \Delta\over \Delta^2-q_0^2-i\epsilon} 
\,.
\eea
$\Delta=m_{\Delta}-m_p$, $b_1^F$ is the 
$\Delta-$photon coupling, for extra details see \cite{Pineda:2002as}. 

We obtain the same result for these structure functions, either 
if we compute them with dispersion relations, or with Feynman 
rules.

\end{document}